\documentclass[a4paper,12pt]{cernyrep}
\pdfoutput=1
\usepackage[utf8]{inputenc}
\usepackage[T1]{fontenc}
\usepackage{a4wide, lmodern, amsmath, amsfonts, graphicx, xcolor, ifthen}
\hypersetup{colorlinks=true, citecolor=blue, urlcolor=blue, linkcolor=blue, breaklinks=true}

\let\Re\undefined
\DeclareMathOperator{\Re}{Re}

\preprint{CERN-LPCC-2022-01\\CERN-LHCEFTWG-2021-002}

\title{LHC EFT WG note:\\Truncation, validity, uncertainties}
\date{\today}

\author{\emph{Editors:}
Ilaria Brivio,\twoaff{Heidelberg}{Zurich}
Sally Dawson,\aff{Brookhaven}
Jorge de Blas,\twoaff{Granada}{CERNTH}
Gauthier Durieux,\aff{CERNTH}
Giovanni Petrucciani,\aff{CERNEP}
Pierre Savard\,\aff{Toronto}
\\[2mm]
\emph{Proposal contributors:}\\[1mm]
\emph{A.}
Roberto Contino,\aff{Sapienza}
Adam Falkowski,\aff{Orsay}
Florian Goertz,\aff{MPIHeidelberg}
Christophe Grojean,\aff{DESY}
Fabio Maltoni,\twoaff{Louvain}{Bologna}
Giuliano Panico,\aff{INFNFirenze}
Francesco Riva,\aff{Geneva}
Andrea Wulzer\aff{Padova}
\\[2mm]
\emph{B.}
Céline Degrande,\aff{Louvain}
Fabio Maltoni,\twoaff{Louvain}{Bologna}
Ken Mimasu,\aff{Kings}
Eleni Vryonidou,\aff{Manchester}
Cen Zhang\threeaff{IHEP}{UCAS}{Peking}
\\[2mm]
\emph{C.}
William Shepherd\,\aff{SamHouston}
\\[2mm]
\emph{D.}
Nicolas Berger,\aff{Annecy}
Andrei V.\ Gritsan,\aff{JHU}
Kristin Lohwasser\aff{Sheffield}
}

\institute{%
\naff{Heidelberg}{Institut für Theoretische Physik, Universität Heidelberg, Germany}
\naff{Zurich}{Physik Institut, University of Zurich, Switzerland}
\naff{Brookhaven}{Department of Physics, Brookhaven National Laboratory, Upton, N.Y., 11973, U.S.A.}
\naff{Granada}{CAFPE and Departamento de Física Teórica y del Cosmos, Universidad de Granada, Campus de Fuentenueva, E–18071 Granada, Spain}
\naff{CERNTH}{Theoretical Physics Department, CERN, 1211 Geneva 23, Switzerland}
\naff{CERNEP}{Experimental Physics Department, CERN, 1211 Geneva 23, Switzerland}
\naff{Toronto}{University of Toronto, Toronto, ON M5S 1A7, Canada}
\naff{Sapienza}{Università di Roma La Sapienza, Piazzale Aldo Moro 5, 00185 Roma, Italy}
\naff{Orsay}{Université Paris-Saclay, CNRS/IN2P3, IJCLab, 91405 Orsay, France}
\naff{MPIHeidelberg}{Max-Planck-Institut für Kernphysik, Saupfercheckweg 1, 69117 Heidelberg, Germany}
\naff{DESY}{DESY, Notkestrasse 85, 22607 Hamburg, Germany}
\naff{Louvain}{Centre for Cosmology, Particle Physics and Phenomenology (CP3), Université catholique de Louvain, Chemin du Cyclotron, B-1348 Louvain la Neuve, Belgium}
\naff{Bologna}{Dipartimento di Fisica e Astronomia, Università di Bologna and INFN, Sezione di Bologna, Via Irnerio 46, 40126 Bologna, Italy}
\naff{INFNFirenze}{INFN Sezione di Firenze and Dipartimento di Fisica e Astronomia, Università di Firenze, Via G.\ Sansone 1, I-50019 Sesto Fiorentino, Italy}
\naff{Geneva}{Départment de Physique Théorique, Université de Genève, 24 quai Ernest-Ansermet, 1211 Genève 4, Switzerland}
\naff{Padova}{Dipartimento di Fisica e Astronomia, Università di Padova, Padova, Italy}
\naff{Kings}{Department of Physics, King's College London, Strand, WC2R 2LS London, UK}
\naff{Manchester}{Department of Physics and Astronomy, University of Manchester, Oxford Road, Manchester M13 9PL, United Kingdom}
\naff{IHEP}{Institute of High Energy Physics, Chinese Academy of Sciences, Beijing 100049, China}
\naff{UCAS}{School of Physical Sciences, University of Chinese Academy of Science, Beijing 100049, China}
\naff{Peking}{Center for High Energy Physics, Peking University, Beijing 100871, China}
\naff{SamHouston}{Department of Physics and Astronomy, Sam Houston State University, Huntsville TX, USA}
\naff{Annecy}{LAPP, Université Savoie Mont Blanc, CNRS/IN2P3, Annecy, France}
\naff{JHU}{Department of Physics and Astronomy, Johns Hopkins University, Baltimore, MD 21218, USA}
\naff{Sheffield}{Department of Physics and Astronomy, University of Sheffield, Sheffield S3 7RH, UK}%
}

\begin{document}
\maketitle

The truncation of the standard-model effective field theory (SMEFT), its validity and the associated uncertainties have been discussed in a dedicated meeting on \href{https://indico.cern.ch/event/980681}{January 19, 2021}.
Answering a call issued beforehand, three proposals were presented:
\href{https://indico.cern.ch/event/980681/contributions/4178323/attachments/2169392/3662359/EFTvalidity.pdf}{\emph{A}},
\href{https://indico.cern.ch/event/980681/contributions/4178322/attachments/2169845/3663208/eft_trunc_unc_val.pdf}{\emph{B}}, and
\href{https://indico.cern.ch/event/980681/contributions/4178321/attachments/2169217/3662006/proposal.pdf}{\emph{C}}.
A preliminary version of the present note summarizing them was written by the editors, submitted for feedback to the proponents, and presented at the \href{https://indico.cern.ch/event/1016713/}{May 3} general meeting.
\href{https://docs.google.com/document/d/13gLoLsELfBaifcTwhSXkcj6z152uz-xlB_WDx2HirFo/edit}{Comments} from the wider community were collected in an online document.
Experimental collaborations provided formal feedback during a second dedicated meeting on \href{https://indico.cern.ch/event/1048848/}{June~28}.
The first version of this summary note was released on January 12, 2022.
Proposal \href{https://cernbox.cern.ch/index.php/s/dnRUj2LPyYNFTVB}{D} was circulated on May 18 and \href{https://docs.google.com/document/d/1lEGgA0oEdqZOEBvyORcReyL0n9ChftQMa-Dd-zv9EQ0}{comments} collected, ahead of the \href{https://indico.cern.ch/event/1136803/}{May 23} general meeting.
Extensive discussions took place with the whole community but no consensus emerged.
None of the proposals has been approved or validated.
No recommendation is therefore put forward at this time and this note only aims at summarizing the different points raised at meetings.
Further work is needed to establish a prescription.
In particular, the benchmarking of the different proposals on the working-group fitting exercise has been \href{https://indico.cern.ch/event/1076709/}{proposed} and \href{https://indico.cern.ch/event/1136803/}{discussed}.

\section{Common ground}
There are various points of agreement between proponents of various schemes for dealing with truncation uncertainties.
Most participants agree that:
\begin{enumerate}

\item most near-future experimental analyses will not aim at probing simultaneously both dimension-six and dimension-eight operator contributions.
The SMEFT truncation of interest is then at the level of dimension-six operators.

\item \label{item:squares} although they only constitute a partial set of $1/\Lambda^4$ corrections, the squares of amplitudes featuring a single dimension-six operator insertion provide a convenient proxy to estimate $1/\Lambda^4$ corrections, as they are well defined and unambiguous. They are indeed gauge invariant and can be translated exactly from one dimension-six operator basis to the other. See \ref{sec:squares} for more detailed statements.

\item \label{item:power-counting} estimating the relative contributions of dimension-six and dimension-eight operators requires a power counting covering a given range of new-physics scenarios and depends on its parameters (e.g.\ mass scales and couplings).
Being able to compute the dimension-eight dependence of observables is insufficient, as a prescription determining the relative magnitude of dimension-six and dimension-eight operator contributions is still needed.

\end{enumerate}

\section{Proposals}
Specific points made in each proposal are succinctly summarized under Arabic numbers.
Additional considerations highlighting pros and cons are listed under Latin lowercases.

Experimental results obtained by following one of the proposals would not be sufficient to allow for the a-posteriori application of the other prescription.

\subsection{Proposals \emph{A} and \emph{B}}
\label{sec:AB}

Proposal \emph{A} and \emph{B} are similar thus discussed together.
They advocate:
\begin{enumerate}

\item providing full multidimensional information on the constrained EFT parameter space, to allow for the proper interpretations in (classes of) new-physics scenarios, and therefore for the EFT validity assessment.
	\begin{enumerate}
	\item Providing individual and marginalized constraints, on single coefficients and in two dimensions, is a first but insufficient step. Full likelihoods, or covariance matrices in Gaussian cases, should be published.
	\item Correlations between operator coefficients deriving from specific new-physics assumptions may exclude parameter-space region where linear and quadratic dimension-six truncations diverge significantly and therefore improve the EFT validity.
	\end{enumerate}

\item including squared dimension-six dependences by default and comparing results with those obtained in the linear SMEFT approximation.
	\begin{enumerate}
	\item The conclusions drawn from this comparison are more qualitative than quantitative.
	
	In case the two sets of results match each other, one can conclude on the general validity of the dimension-six truncation, as situations in which dimension-eight contributions would dominate over linear dimension-six ones are likely pathological. No additional assumption is required on new physics.
	
	When the linear and quadratic results differ significantly, constraints can only be applied in scenarios where dimension-eight contributions are generically suppressed with respect to quadratic dimension-six ones.
	
	\item The linear-quadratic comparison does not reflect the convergence of the EFT series when interference contributions suffer (accidental or understood) suppressions.
	
	\item The assignment of a truncation uncertainty is not prescribed.
	
	\item \gdef\linearfits{Purely linear fits can be technically more involved, as they formally allow event rates to turn negative.}\linearfits

	\end{enumerate}

\item\label{item:clipping} providing experimental results as functions of the maximal energy probed in the data employed, introducing where necessary an upper cut (denoted e.g.\ $E_\text{cut}$ or $M_\text{cut}$).
Data and prediction are compared in the same phase-space region.
This procedure, often referred to as \emph{clipping}, provides the necessary information to verify the EFT validity for (classes of) models and enhances the range of scenarios for which a valid EFT interpretation is possible.

	\begin{enumerate}
	\item For different cut values, different analysis strategies may be required.
While rate measurements could provide the highest sensitivities at high energies, differential observables may be required to probe the relevant operators at moderate energies.
Upper energy cuts should therefore be considered from the onset in the analysis design.

	\item\label{item:clipping-complication} The reconstruction of the relevant variable to cut on may complicate experimental analyses and result in additional systematic uncertainties (e.g.\ in final states featuring missing energy).
Many EFT analyses (e.g.\ STXS in the Higgs sector) do however already measure suitable energy variables (e.g.\ bosons' transverse momenta, or jet invariant masses) as sensitivity arises from these.
Example of experimental analyses having adopted such a clipping procedure include diboson measurements by CMS (in \href{https://cms-results.web.cern.ch/cms-results/public-results/preliminary-results/SMP-20-005}{$W\gamma$},
\href{https://cms-results.web.cern.ch/cms-results/public-results/preliminary-results/SMP-20-014}{$WZ$},
\href{https://cms-results.web.cern.ch/cms-results/public-results/publications/SMP-16-017}{$ZZ$} final states).

	\item\label{item:clipping-combination} Combining different observables from different processes, each using an upper cut on a different variable, may also raise questions.
What variables and cut values are compatible in different processes?
The study of specific scenarios may be informative in that regard.
Conclusions are expected to be model dependent.

	\item\label{item:clipping-cost} Repeating global analyses for several upper cut values would be more costly both computationally and in term of personpower.
	
	\item\label{item:sim-clipping} Applying a cut on the EFT signal simulation instead of the data was proposed by experimental collaborations, as described later in \autoref{sec:proposal-D}.
As existing run-2 analyses will not be re-designed, modifying only the signal simulation could in particular be used to incorporate, into EFT fits, analyses in which no good energy variable was measured.
The proponents \emph{A} and \emph{B} however judge that comparing data in a given phase-space region (without energy cut) with predictions in a different one (with energy cut) is inconsistent.
Further studies could clarify whether the two approaches are practically equivalent in cases of interest.

	\end{enumerate}

\item assessing, a posteriori (even after the combination of different measurements), the range of models for which the extracted constraints apply, using this $E_\text{cut}$ information.
The experimental results themselves would therefore not incorporate assumptions about new-physics models.
This approach allows theorists to interpret the results in the context of specific (classes of) models.
	\begin{enumerate}
	\item Quantifying missing dimension-eight contributions would require more effort.
	
	A posteriori, one could approximately reproduce the experimental analysis and include estimates of dimension-eight contributions in the relevant phase-space region to assess their impact.
	
	A priori, one could consider treating linear and quadratic contributions as independent in experimental analyses and, in interpretations, rescale the quadratic contribution to estimate dimension-eight effects. The significant increase in the number of parameters to be fitted may however not be practical.
	
	\end{enumerate}

\end{enumerate}

\subsection{Proposal \emph{C}}
Proposal \emph{C} advocates:
\begin{enumerate}

\item using squared dimension-six contributions, which can readily be computed with existing tools, as proxies for missing dimension-eight terms at order $1/\Lambda^4$.
	\begin{enumerate}
	\item As the dimension-eight contributions at that order arise from interferences with SM amplitudes, the dimension-eight contributions could have different kinematic distributions or suffer some accidental suppressions.
	
	\item It is claimed (see online \href{https://docs.google.com/document/d/13gLoLsELfBaifcTwhSXkcj6z152uz-xlB_WDx2HirFo/edit}{comments}) that such contributions violate gauge invariance, in contradiction with \autoref{item:squares} and \ref{sec:squares}.
	\end{enumerate}

\item employing a power-counting rule that would encompass many new-physics models, to estimate dimension-eight contributions from squared dimension-six ones.
	\begin{enumerate}
	\item The line drawn between classes of models that are, and are not, covered by the chosen power-counting rule is somewhat arbitrary.
	\item For specific classes of scenarios, this will necessarily be overly conservative. Employing different power-counting rules for different classes of scenarios would permit to quote tighter constraints in the specific cases where they apply.
	\end{enumerate}

\item considering the known squared dimension-six terms together with the dimension-eight estimates as theoretical uncertainty on the linear dimension-six signal.
	\begin{enumerate}
	\item As the dimension-six squared contributions are known, they may not need to be  included in uncertainties.
	\item This uncertainty depends on the SMEFT parameter point and could therefore be practically difficult to include in analyses.
	\item \linearfits
	\end{enumerate}

\item folding these uncertainties directly into experimental analyses.
	\begin{enumerate}
	\item This renders experimental results model dependent, as they then rely on a specific scaling between dimension-six and dimension-eight operator coefficients.
	\end{enumerate}

\item multiplying the squared dimension-six terms by the following factor to obtain truncation uncertainties:
	\begin{equation}
	1 + \sqrt{N_8} \frac{g_\text{SM}^2}{\mathfrak{C}_6\Lambda^2} \sqrt{1+\frac{1}{\mathfrak{C}_6^2\Lambda^4}}
	\end{equation}
	where $N_8$ is an estimate of the number of contributing dimension-eight operators, $g_\text{SM}$ is the relevant SM coupling, $\mathfrak{C}_6$ is a dimensionful dimension-six operator coefficient, and $\Lambda$ is a scale (such that, if identified with the physical BSM mass scale in a two-to-two process, one expects $\mathfrak{C}_6\Lambda^2\sim g_\text{BSM}^2$).
	\begin{enumerate}
	\item The classes of models which are covered by this choice is yet to be determined.
	Without motivation from new-physics models, the various factors may seem ad hoc.
	\end{enumerate}

\item in cases where dimension-eight contributions can be computed, the functional form of dimension-six contributions squared would not need to be used as proxy for the dimension-eight ones and a power counting could be used for dimension-eight (and dimension-six) coefficients directly.

\end{enumerate}

\subsection{Proposal D}
\label{sec:proposal-D}

The experimental proposal \emph{D} suggests to clip the EFT simulation, instead of clipping the data as prescribed by theorists in \autoref{item:clipping} of \autoref{sec:AB}.
It stressed that:

\begin{enumerate}

\item redesigning existing analyses in cases where no energy variable suitable to perform data clipping was measured is difficult (\autoref{item:clipping-complication}) and, where the final fitted variable is not an energy one (e.g.\ if multivariate techniques are used to maximise EFT sensitivity), re-running analyses for several energy cut values is costly (\autoref{item:clipping-cost}).
	
	\begin{enumerate}
	\item Presently, many relevant (e.g.\ unfolded) analyses measure the distribution of an energy variable from which bins could straightforwardly be removed when performing global EFT fits (\autoref{item:clipping-complication}).

	\end{enumerate}

\item clipping the EFT simulation in the tail of an energy variable yields indications about the impact of this kinematic region on extracted EFT parameters.
Proper constraints are derived using the full dataset and simulation, and a comparison with results obtained with simulation clipping provides qualitative information about the sensitivity to high energies.

	\begin{enumerate}
	\item Data clipping, by confronting data and simulation in the same phase-space region, additionally allows to derive proper constraints for models with characteristic energy scales lower than the energies directly probed in the considered measurement (\autoref{item:clipping}).
	\end{enumerate}

\item the EFT simulation can be clipped on any desired variable and at any value, at the cost of a re-computation of the EFT prediction, while clipping the data is only possible on measured variables and at bin boundaries (without re-running the analysis).

	\begin{enumerate}
	\item There is however no universal best choice of clipping variable.
Different models can induce energy growths in different kinematic quantities (e.g.\ $s$ and $t$ channel mediators in the Drell-Yan process, both leading to four-fermion operators).
Different operators contributing to the same process could also induce energy growth in different variables.

	\item As for data clipping, the combination of different processes clipped on different variables, at different values, is an open question (\autoref{item:clipping-combination}).
Explicit simplified models could be examined to inform such choices.
One could explore the possibility of using quantile of the SM distribution in each energy variable.
For data clipping, a public fitting framework with high flexibility in the bins of data included could allow the community to adopt the choices best suited for any particular interpretation.

	\end{enumerate}

\item the clippings of data and EFT simulation are equivalent if performed on a variable that is actually measured.
For data clipping at the reconstructed level and simulation clipping at the truth Monte Carlo level, the two procedures would only differ by finite resolution effects.
If the data does not match the SM expectation, a poor fit would be obtained.
	
	\begin{enumerate}
	\item On the other hand, clipping the EFT simulation on a variable that is not measured could lead to pathologies arising from the mismatch between the phase spaces of the data and prediction (\autoref{item:sim-clipping}).
Correlations between the clipped and measured variables could then lead to an overestimation of the strength of the constraints derived when the data agrees with the SM.
In case the data deviates from the SM, the extracted operator coefficients could also be biased away from their true value, in a process- and observable-dependent fashion.

	\end{enumerate}

\end{enumerate}

\appendix
\renewcommand{\thesection}{Appendix A}
\section{Well-defined and unambiguous squares}
\label{sec:squares}

In the $1/\Lambda^2$ expansion of dimension-six amplitudes ($S$-matrix elements), the zeroth and first terms, $A_\text{SM}$ and $A_6/\Lambda^2$, are separately gauge invariant.
The $A_6/\Lambda^2$ term contains all and only $1/\Lambda^2$ contributions to the amplitude.
The squares of those two terms ---$|A_\text{SM}|^2$, $\Re\{A_\text{SM}^* A_6\}/\Lambda^2$, $|A_6|^2/\Lambda^4$--- are thus separately gauge invariant too.
More terms do appear in the squared amplitude at order $1/\Lambda^4$: from dimension-eight operators, amplitudes with two dimension-six operator insertions, or field redefinitions expanded to $1/\Lambda^4$ order.
Each of these other subclasses of $1/\Lambda^4$ contributions to the squared amplitude is in general not separately gauge invariant: only the full $\Re\{A_\text{SM}^*A_8\}/\Lambda^4$ is, where $A_8/\Lambda^4$ collects all and only $1/\Lambda^4$ contributions to the amplitude and is itself separately gauge invariant.

As a consequence of the equivalence theorem, dimension-six operators related by classical equations of motion have identical amplitudes ($S$-matrix elements) to $1/\Lambda^2$ order.
Equivalent operator bases can be defined by exploiting this freedom, changing operator normalizations, or taking linear combinations of them.
The linear dimension-six amplitude $A_6/\Lambda^2$, including all and only $1/\Lambda^2$ contributions, can thus be translated exactly from one dimension-six operator basis to the other by a linear transformation between the two sets of operator coefficients $\{c_6\}$ and $\{c'_6\}$.
The same transformation can also be used to translate exactly the square of this linear dimension-six amplitude $|A_6|^2/\Lambda^4$ from one dimension-six basis to the other.

For these two specific reasons, the square of the linear dimension-six amplitude $|A_6|^2/\Lambda^4$, where $A_6/\Lambda^2$ contains all and only $1/\Lambda^2$ contributions to the amplitude, can be qualified as \emph{well-defined} and \emph{unambiguous}.
It can thus for instance be employed as a convenient proxy for estimating full $1/\Lambda^4$ contributions.

\bigskip\noindent\textbf{Acknowledgments}:
This summary note has been written under the umbrella of the LHC EFT WG and we would like to thank all the participants who contributed to the corresponding discussions.

\end{document}